\documentclass[sigconf]{acmart}
\usepackage{amssymb,amsmath,algorithm,algorithmic,graphicx,subfigure}

\newtheorem{prop}{Proposition}

\newcommand\ep{\varepsilon}
\newcommand\R{\mathbb{R}}

\newif\ifformatforarxiv
\formatforarxivtrue
\ifformatforarxiv
  \renewcommand\footnotetextcopyrightpermission[1]{} 
\fi

\begin{document}
\title{SCRank: Spammer and Celebrity Ranking in Directed Social Networks}
\author{Alex Fabrikant}
\affiliation{%
\institution{Google Research}
\city{Mountain View}
\state{CA}
\country{USA}}
\email{fabrikant@google.com}
\author{Mohammad Mahdian}
\affiliation{%
\institution{Google Research}
\city{New York}
\state{NY}
\country{USA}}
\email{mahdian@google.com}
\author{Andrew Tomkins}
\affiliation{%
\institution{Google Research}
\city{Mountain View}
\state{CA}
\country{USA}}
\email{tomkins@google.com}

\begin{abstract}
Many online social networks allow directed edges: Alice can unilaterally add an ``edge'' to Bob, typically indicating some kind of interest in Bob, or in Bob's content, without Bob necessarily reciprocating with an ``add-back'' edge that would have indicated Bob's interest in Alice. This significantly affects the dynamics of interactions in the social network. Most importantly, we observe the rise of two distinctive classes of users, {\em celebrities} and {\em follow spammers}, who accrue unreciprocated directed links in two different directions: celebrities attract many unreciprocated incoming links, and follow spammers create many unreciprocated outgoing links.  Identifying users in both of these two categories is an important problem since a user's status as a celebrity or as a follow spammer is an important factor in abuse detection, user and content ranking, privacy choices, and other social network features.

In this paper we develop SCRank, an iterative algorithm that exploits a deep connection between these two categories, and classifies both celebrities and follow spammers using purely the social graph structure. We analyze SCRank both theoretically and experimentally. Our theoretical analysis shows that SCRank always decreases a potential function, and therefore converges to an approximate equilibrium point. We then use experimental evaluation on a real global-scale social network and on synthetically generated graphs to observe that the algorithm converges very quickly, and consistently to the same solution.  Using synthetic data with built-in ground truth, we also experimentally show that the algorithm provides a good approximation to the built-in set of celebrities and spammers.  Finally, we generalize our convergence proof to a general class of ``scoring'' algorithms, and prove that under mild conditions, algorithms in this class minimize a (non-trivial) potential function and therefore converge.  We give several examples to demonstrate the versatility of this general framework and usefulness of our techniques in proving theoretical results on the convergence of iterative algorithms.
\end{abstract}

\maketitle

\section{Introduction}
\label{sec:intro}

Online social networks can be divided into two categories: undirected
networks such as LinkedIn or (pre-2011) Facebook that require the
consent of both endpoints in order to establish an edge, and directed
ones such as Twitter, Google$+$, and Flickr, that allow one user to
unilaterally create a directed edge to another, such as by
``following'' the latter's public updates, without the latter creating
a reciprocal (anti-parallel) edge to the former. As has been
observed in practice \cite{OReillyRant},
this simple distinction significantly affects
the dynamics of relationships in the system: undirected social
networks like Facebook tend to cultivate socializing with friends,
while directed networks like Twitter, interacting with content produced by
strangers constitutes a significant portion of social interactions.
In the latter case, there are often a few individuals who collect many
incoming links, either because they are already famous outside the
social network, or because they contribute exceptionally engaging,
viral content to the social network's ecosystem. These individuals are
in a sense the ``celebrities'' of the network. On the other hand, we
have nodes who accumulate many outgoing links to random strangers. We
call these nodes ``(follow) spammers''.  As we explain below, identifying
celebrities and spammers of a network are intertwined
problems. This paper focuses on developing algorithms to identify
users in these two classes.

The simplest approach to identify a spammer is to count the number of
unreciprocated outgoing edges of each node and classify the node as a
spammer if this number exceeds a threshold.  The problem with this
approach is that often a non-negligible number of regular users follow
many celebrities, and this approach can identify such users as
spammers.  Similarly, classifying celebrities by counting the number
of unreciprocated incoming links suffers from the problem that it can
classify regular users who are targeted by many spammers (for example,
by the virtue of having their name mentioned in a public, crawlable
space) as celebrities.  Instead, we focus on this recursive 
definition of celebrities and spammers:

\begin{itemize}
\item A celebrity is a node who is followed by many non-spammers.
\item A spammer is a node who follows many non-celebrities.
\end{itemize}

This recursive definition hints at a natural iterative approach for
finding celebrities and spammers. In the next section, we
mathematically formulate the problem and the iterative algorithm,
which we call SCRank. We then analyze the convergence properties of
SCRank, both theoretically and experimentally, and argue that its
output provides useful information. We will use a real-world data set
from LiveJournal, as well as randomly generated data, to
experimentally evalute the convergence properties of our algorithm.
To evaluate the output of our algorithm, we use randomly generated
data with built in ground-truth, and show that the algorithm can
recover a significant portion of the ground truth efficiently and
accurately.

Finally, in Section~\ref{sec:gen}, we give a generalization of our
potential function argument to a more general framework of scoring
problems, and prove that for any scoring problem satisfying a mild
symmetry and monotonicity assumption, a (non-trivial) potential
function can be associated with the natural iterative algorithm for
the problem, and therefore the iterative algorithm provably
converges. We give three concrete examples of this general framework
to demonstrate the versatility of our framework.

\subsection{Related work}

Various measures of an individual's standing in a social network has
been the subject of much research in sociology and social computing,
starting well before the dawn of online social networks.

Among the two axes we study, celebrities and follow spammers, the
lion's share of the prior work on social graph structure has focused
on celebrities, typically with a goal of understanding and
algorithmically locating highly influential people for the purposes of
ranking, marketing, predicting cascades, etc. \cite{Freemann77},
\cite{Bonacich87}, and many other early sociometric studies focused on
defining and evaluating social centrality metrics. In the digital age,
algorithms for selecting high-influence sets of social network users
from the social graph structure were pioneered by \cite{KKT03},
followed by a large literature of its own. Much of the search engine
literature focuses on finding influential nodes on the
web graph, with the results on PageRank
\cite{PageRank} and HITS \cite{HA} forming perhaps the most
influential nodes in the citation network. These and related
techniques have been borrowed for social network applications as well,
such as by \cite{WhiteSmyth2003}. While much of this work has focused
on the relatively more sophisticated notion of influence, as measured
by impact on viral cascades, less attention has been paid to
questioning the idea that a high in-degree determines a user's
``celebrity'' status. For the corresponding problem on the Web graph,
\cite{Upstill2003} notably gave experimental evidence that corporate
websites' in-degree is a better predictor of a company's prominence
and worth than PageRank.

The follow spam problem has been recognized for several years now
\cite{GawkerTwitter, Ghosh12}, but most of the existing work that does consider
the structure of the social graph still focuses on holistic
machine-learning approaches that combine graph properties with a many
signals derived from user content \cite{Wang10, benevenuto2009, SLK2011}
--- a very pragmatic approach for detecting existing spamming
activity, but of limited utility in the common case where creating
sibyl accounts is cheap \cite{Yang2011}, and most abuser accounts are
thus young.

The existing approaches also assign some form of ``trust'' semantics
to each directed edge, typically making it difficult to cope with
``social capitalists'' \cite{Ghosh12}: the many legitimately popular
celebrities such as Barack Obama or Lady Gaga who have been observed
to reciprocate follow edges indiscriminately. Even when such
indiscriminate behavior is fairly common, SCRank is unaffected, since
it entirely ignores reciprocated edges, and requires only a fraction
of users to be discerning about follow-backs to get enough input
signals.

The potential function that makes our analysis work combines the
potential functions of potential games \cite{potentialgames} and
Max-Cut games \cite{maxcutgame}. The form of the SCRank algorithm
itself is inspired various iterative numerical algorithms used in
machine learning such as EM and belief propagation \cite{MLBook}, and
more directly by the HITS \cite{HA} algorithm for web ranking. The
differences between SCRank and HITS are subtle, but vital to
understanding the operation and analysis behind SCRank, so we now
address this specific relationship.

\subsection{SCRank vs HITS}

At first blush, our reciprocal definition of spammers and celebrities
in terms of one another appears parallel to the definitions of hubs
and authorities in HITS. But mathematically, the structures are
quite different.  HITS is expressible as a linear transformation of
either the original hub or authority vector, which converges by the
Perron-Frobenius theorem to the (positive real) principal eigenvector
of a matrix based on the original graph. There are two properties
that set the spammer-celebrity iteration apart from HITS.  First, the
core update step of the spammer-celebrity iteration involves an affine
transformation rather than a linear transformation; such
transformations do not in general attain fixed points.  Thus, we use
the current spammer vector $\vec{s}$ to compute an intermedate
celebrity value $\vec{c'} = A(\textbf{1}-\vec{s})$ via an affine
transformation.  Second, for reasons we describe below, the particular
update we seek requires an elementwise modification to the results of
the affine transformation by an arbitrary increasing function $F_s$.
The new version of $\vec{s}$ is given by $\vec{s} = F_s(\vec{c'})$.
The actual transformation is therefore no longer affine, but will in
general be non-linear. Likewise, a similar transformation applies to
produce a new spammer vector from the current celebrity vector.
Combining both the affine transformation and the non-linear
modification, we have $\vec{c} =
F_c\left(A^t(\textbf{1}-\vec{s})\right)$.

We now offer two words of intuition on the form of our update.  First,
the affine structure comes about because, unlike HITS, outliers on the
celebrity scale provide no information about spammers.  On the
contrary, only nodes that receive low scores on one scale may provide
significant contributions to the score of nodes on the other scale.  A
little algebraic manipulation will convince the reader that this
property is fundamental to the nature of the relationship between
these classes, and cannot be overcome by simple linear transformations
of the variables, such as introducing ``non-spammer'' scores or the
like.

Second, the non-linear transfer function comes about for a related
reason.  The $\textbf{1}-\vec{s}$ term can be interpreted as a
``non-spammer'' score if spammer scores lie in $[0,1],$ but in general
if spammer scores may grow large, the affine transformation will
produce large negative non-spammer scores, which break the intuition
that links from spammers should not contribute one way or the other to
celebrity scores.  Thus, scores must be scaled to remain in $[0,1]$ in
order to perform the iteration with the semantics we desire.

In general, the	machinery we develop here is appropriate in any
situation that shows anti-reinforcing behavior: shady groups fund dishonest
politicians, while honest politicians are funded by non-shady groups;
and so forth.

\section{The Algorithm}
\label{sec:alg}

To formalize an algorithm based on the recursive definition of the
celebrities and spammers in the previous section, we
define a {\em celebrity score} $c_v$ and a {\em spammer score} $s_v$
for each node $v$. All these scores are in $[0,1]$. The algorithm is
parameterized by two increasing functions $F_c$ and $F_s$ that map
non-negative reals to $[0,1]$. We denote the directed social network
by $G$, and the vertex set, the edge set, and the number of vertices
of $G$ by $V$, $E$, and $n$. Also, the set of {\em unreciprocated} directed
edges of $G$ is denoted by $A$. In other words, 
$A=\{(u,v): (u,v)\in E \mbox{ and } (v,u)\not\in E\}$.

The algorithm in presented in detail as Algorithm~\ref{alg1}. We refer
to this algorithm as the {\em SCRank} algorithm, for Spammer-Celebrity Rank.
The algorithm is based
on iterating the following two assignments until either an approximate fixed point
is found, or a maximum number of iterations is reached:

\begin{equation}
\label{eqn:main}
c_v = \displaystyle F_c \left(\sum_{(u,v)\in A}\left( 1-s_u\right)
\right)
\quad
s_v = \displaystyle F_s \left(\sum_{(v,u)\in A}\left( 1-c_u\right) \right)
\end{equation}

\begin{algorithm}
\caption{\bf (SCRank)}
\label{alg1}
\begin{algorithmic}[1]
\REQUIRE Directed social network $G$; functions $F_c$ and $F_s$;\\
parameters $\ep$ and $T$

\ENSURE Celebrity score $c_v$ and spammer score $s_v$ for each $v\in                                                                                                                       
V$\\[3mm]

\FORALL{$v$ in $V$}
\STATE $c_v \leftarrow 0$, $s_v\leftarrow 0$
\ENDFOR
\STATE $k\leftarrow 0$
\REPEAT
\FORALL{$v$ in $V$}
\STATE $c^{new}_v \leftarrow F_c(\sum_{(u,v)\in A}(1-s_u))$
\ENDFOR
\FORALL{$v$ in $V$}
\STATE $s^{new}_v \leftarrow F_s(\sum_{(v,u)\in A}(1-c^{new}_u))$
\ENDFOR
\STATE $\delta \leftarrow \max\{||c - c^{new}||_\infty, ||s - s^{new}||_\infty\}$
\STATE $c\leftarrow c^{new}$
\STATE $s\leftarrow s^{new}$
\STATE $k\leftarrow k+1$
\UNTIL $\delta < \ep$ or $k > T$
\end{algorithmic}
\end{algorithm}

For our experiments, we will use the CDF of a normal distribution with
mean $\mu_c$ and standard deviation $\sigma_c$ as the function
$F_c$. This is essentially a soft step function where $\mu_c$ controls
the location of the step (the threshold for the number of non-spammer
followers to count a user as a celebrity) and $\sigma_c$ controls the
smoothness of this step function (a large $\sigma_c$ means a smooth
threshold at $\mu_c$, while a small $\sigma_c$ we get a sharp
threshold). 
Similarly, we use the CDF of normal
distribution with mean $\mu_s$ and standard deviation $\sigma_s$ as
$F_s$.

We note that our results go through even if the functions $F_s$ and
$F_c$ depend on the vertex $v$. This might be practically useful,
for example, by allowing the threshold $\mu_s$ to depend on the number
of reciprocated neighbors of the vertex (i.e., if a node has a large
number of reciprocated edges, allow it to have more unreciprocated
edges without counting it as a spammer). This and further
generalizations will be discussed in Section~\ref{sec:gen}.

Our algorithm is similar in spirit to the Hubs and Authorities
algorithm of Kleinberg~\cite{HA}. The major difference is that
in our setting, the celebrity score of a node is related to the 
{\em non}-spammer score of its followers. This negation makes a
significant difference: we need the spammer scores to be scaled in
$[0,1]$ with 0 meaning a non-spammer and 1 meaning a spammer (and
similarly for celebrities), whereas in the hubs and authorities
algorithm it was enough to compute scores that induce reasonable
rankings. This, forces us to use non-linear operators $F_s$
and $F_c$. This is in contrast with hubs and authorities, which uses
linear operators and therefore can characterize the scores as
eigenvectors of a matrix.

\paragraph{Note on the implementation}
In order to be able to use the SCRank algorithm on graphs with
hundreds of millions of nodes (as we do in Section~\ref{sec:exp}), we
need to take advantage of parallel computation. Fortunately, for the
SCRank algorithm this is not hard to do, since the celebrity scores in
each iteration only depend on the spammer scores last computed and
vice versa. Using this, we implemented each iteration of SCRank as two
Map-Reduce stages, without any blow-up in the size of the data in each
iteration. This yielded a very efficient implementation which easily
accommodated even our largest experiments in Section~\ref{sec:exp}, on
a social graph of over 400,000,000 nodes.

\section{Convergence of the algorithm}
\label{sec:conv}
Ideally, we would like to show that: (1) when there is no bound $T$ on
the number of iterations, SCRank converges to an (approximate)
fixed point; (2) the fixed point is unique; and (3) the algorithm
converges quickly to the fixed point.  In this section, we
theoretically show that (1) holds for all directed social networks.
We will give an example that shows that the fixed point of the
function is {\em not} necessarily unique, much like in HITS and other similar algorithms \cite{hits-nonunique}. However, as we will discuss
in the next section, we have not observed such examples in real or
randomly generated data sets.  Finally, we will experimentally show
that the SCRank algorithm often converges very quickly.

We start by proving that the algorithm never falls into a loop. This is
done by showing that there is a potential function whose value
decreases in every iteration. The intuition behind this
(complicated-looking) potential function is that it combines the
potential function for max cut games~\cite{maxcutgame} with those of
potential games~\cite{potentialgames}.

In the following theorem, we show the existence of this potential
function. We will then use this result to prove that the algorithm
converges to an approximate fixed point of the
Equations~\eqref{eqn:main}.

\begin{theorem}
\label{thm:noloop}
For every directed social network $G$ and every pair of increasing
differentiable functions $F_s$ and $F_c$, there is a function of the
the vector $(c,s)$ computed by the SCRank algorithm that strictly 
decreases in every iteration. Therefore, the algorithm will never fall 
in an infinite loop.
\end{theorem}
\begin{proof}
Let $R_c = F_c([0,n])$, i.e., $R_c$ is the range of $F_c$ when its
domain is $[0,n]$. Since $F_c$ is increasing and differentiable,
its inverse on $R_c$ is a well defined {\em strictly} increasing function
${F_c}^{-1}$. Next, we define the following function:

$$G_c(x) = \int_0^x F_c^{-1}(t)dt.$$

Similarly, using $F_s$, we can define $R_s$ and $G_s$. We
are now ready to define our potential function. For any vector of
celebrity and spammer scores $(c,s)\in\R^{2n}$, the function $P$ is
defined as follows:

\begin{equation}
P(c,s):=\sum_{(u,v)\in A} (1-s_u)(1-c_v) + \sum_{v\in V} G_c(c_v) + \sum_{v\in V} G_s(s_v)
\end{equation}

Next, we show that the value of this potential function decreases in
every iteration of the algorithm. To do this, take the derivative of
$P$ with respect to $c_v$, for a vertex $v$:

$$                                                                                                                                                                                         
\frac{\partial P(c,s)}{\partial c_v} = -\sum_{u: (u,v)\in A} (1-s_u) + F_c^{-1}(c_v)                                                                                                       
$$

This derivative is zero when $c_v$ is equal to $$c^*_v := F_c(\sum_{u: 
 (u,v)\in A}(1-s_u)),$$ negative when $c_v < c^*_v$, and positive when
$c_v > c^*_v$. Therefore, by changing $c_v$ from its old value
$c^{old}_v$ to $c^*_v$, the value of the potential function can not
increase. This means that the updates in line 7 of
Algorithm~\ref{alg1} never increase the value of $P$. In fact, since
$F_c^{-1}$ is a strictly increasing function, if at least one of the
$c_v$'s change, then the potential function must strictly decrease. A
similar argument shows that the updates in line 10 of
Algorithm~\ref{alg1} also do not increase the value of $P$. This is
enough to show that the algorithm never falls into an infinite loop.
\end{proof}

Next, we prove that the SCRank algorithm eventually converges to 
an approximate fixed point (also referred to as an approximate equilibrium).
Before stating the theorem, we need to define the notion of
approximate fixed point.

\begin{definition}
An $\ep$-approximate fixed point of Equations~\eqref{eqn:main} is a set of 
celebrity and spammer scores $(c_v, s_v)$ for each node such that for
each vertex $v$, we have
\begin{equation}
\label{eqn:defnapprox}
\begin{array}{lll}
|c_v - \displaystyle F_c(\sum_{(u,v)\in A}(1-s_u))| & \le & \ep\\[7mm]
|s_v - \displaystyle F_s(\sum_{(v,u)\in A}(1-c_u))| & \le & \ep
\end{array}\end{equation}
\end{definition}

\begin{theorem}
\label{thm:conv}
For every $\ep > 0$, there is a finite number of iterations after which
the vector $(c,s)$ computed by the SCRank algorithm is an
$\ep$-approximate fixed point of Equations~\eqref{eqn:main}.
\end{theorem}

\begin{proof}
We use the notation from the proof of Theorem~\ref{thm:noloop}.
Since $F_c$ is increasing and differentiable on a closed interval $[0,n]$, there
is an absolute constant $\alpha$, such that for every non-negative
$x\in[0,n]$, the derivative of $F_c$ at $x$ is at most
$\alpha_c$. This implies that the derivative of the function
$F_c^{-1}$ on every point in $R_c$ is at least
$1/\alpha_c$. Similarly, we can define $\alpha_s$ for $F_s$ and show
that the derivative of $F_s^{-1}$ on $R_s$ is at least
$1/\alpha_s$. Let $\alpha = \max(\alpha_c, \alpha_s)$.

Next, we prove that if an update operation changes the values by too
much, it must also significantly decrease the value of the potential
function. Assume in an iteration the value of $c_v$ is changed from
$c^{old}_v$ to $c^*_v$, where $|c^{old}_v - c^*_v|\ge\ep$.  Assume
$c^*_v < c^{old}_v$ (the $c^*_v > c^{old}_v$ case can be handled
similarly). Then for every $x\in [c^*_v, c^{old}_v]$, we have

\begin{eqnarray*}
F_c^{-1}(x) &>& F_c^{-1}(c^*_v) + (x-c^*_v)/\alpha_c \\
&=& \sum_{u: (u,v)\in A} (1-s_u) + (x-c^*_v)/\alpha_c.
\end{eqnarray*}

Therefore, the derivative of the function $P(c,v)$ with respect to
$c_v$ at $c_v = x$ is at least $(x-c^*_v)/\alpha_c$. Thus, the value
of $P(c,v)$ at $c_v = c^{old}_v$ is at least its value at $c_v =                                                                                                                           
c^*_v$ plus $\int_{c^*_v}^{c^{old}_v}(x-c^*_v)/\alpha_c dx \ge                                                                                                                             
\frac{\ep^2}{2\alpha_c}$.  In other words, in each iteration where the
value of at least one $c_v$ changes by at least $\ep$, the value of
the potential function decreases by at least
$\ep^2/(2\alpha_c)\ge \ep^2/(2\alpha)$. Similarly, if the value of at 
least one $s_v$ changes by at least $\ep$, the potential function 
decreases by at least $\ep^2/(2\alpha_c)\ge \ep^2/(2\alpha)$. 
Since the value of the potential function decreases in every iteration and can
never become negative, after a finite number of iterations it must
decrease by an amount less than $\ep^2/(2\alpha)$. This means that
at this iteration, each score changes by at most $\ep$, showing that
we are at an $\ep$-approximate fixed point.
\end{proof}

\paragraph{Uniqueness of the fixed point}
It would be nice if we could prove that the fixed point of
Equations~\eqref{eqn:main} is unique. This would mean that the values
that the SCRank algorithm seeks to compute are uniquely well-defined.
Unfortunately, this result is not true in the worst-case, as the
following example shows.

\begin{figure}
$$\includegraphics[width=7cm]{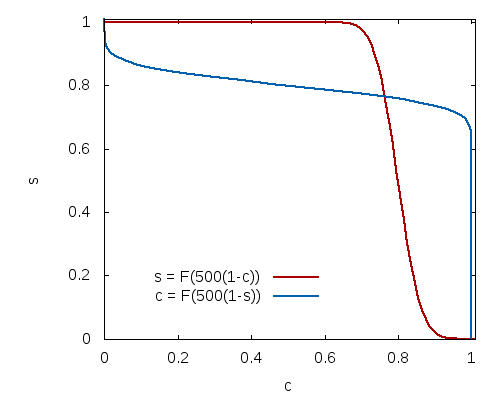}$$
\caption{Example with more than one fixed point}
\label{fig:nonunique}
\end{figure}

\begin{prop}
There is a directed social network $D$ and functions $F_c$ and $F_s$
such that more than one $(c,s)$ satisfies the Equations~\eqref{eqn:main}.
\end{prop}
\begin{proof}
Consider a regular bipartite graph with all the edges directed from
part 1 to part 2. Intuitively, this situation can be explained by
either declaring nodes in part 1 as spammers, or nodes in part 2 as
celebrities.  For a numeric example, say the degrees are 500,
$\mu_s=\mu_c=100$, and $\sigma_s=\sigma_c=25$. Let $F=F_c=F_s$.  Then
nodes in part 1 will have celebrity score $F(0)\approx 0$ and spammer
score $s$, and nodes in part 2 will have celebrity score $c$ and
spammer score $F(0)\approx 0$, for values of $(c,s)$ satisfying $c =
F(500(1-s))$ and $s=F(500(1-c))$. These equations are plotted in
Figure~\ref{fig:nonunique}.  As can be seen in the picture, there are
3 fixed points with $(c,s)$ approximately equal to $(1,0)$, $(0,1)$,
and $(0.8,0.8)$.  The first fixed point corresponds to declaring nodes
in part 1 as spammers, the second corresponds to declaring nodes in
part 2 as celebrities, and the third is an unstable fixed point
between the other two.
\end{proof}

Despite the above example, as we will see in the next section, in none of
the real world or randomly generated instances we have tried we have
been able to discover more than one solution.

\section{Experiments}
\label{sec:exp}
In this section, we present the results of experiments showing that on
real and generated data, the algorithm presented in the last section
converges quickly and to the same point, independent of the starting
configuration. We also show that the computed scores are reasonable
quantifications of celebritiness and spamminess in the social
networks. This is done with randomly generated data sets with a 
random generation process that embeds the ground truth against which
the output of the algorithm can be evaluated. For experimentally
evaluating convergence and uniqueness properties, we use randomly 
generated data as well as two real-world data sets, as described in
the following.

\subsection{Data Sets}
We use two sources of data in our experiments. The first is based on
real-world data from LiveJournal. The second is randomly generated
data according to a model described below. Randomly generated data
allows us to compare the results of the algorithm with the ``ground
truth'' that the model is based on. This is in contrast with the
real-world data set, which is used to evaluate the convergence and
uniqueness properties of the SCRank algorithm. As we show, manually
skimming the results on this data suggests that the outputs are
reasonable, but we do not have quantifiable ground truth.

In the rest of this section, we describe the random generation process
and basic information about our real-world data set.

\paragraph{The random generation process}
We use the following method to generate a random directed graph $G$
that will be used as a test case for our algorithm: There are $N$
nodes in the graph, out of which two disjoint sets $C$ and $S$ are
designated as the set of celebrities and spammers. We then use a graph
generation method such as Erd\H{o}s-R\'enyi or preferential attachment
to generate an undirected graph $H$ with the vertex set $V(G)$. The
edges of this graph represent real friendship relationships among
individuals. For each such edge $uv$ in $H$, with probability $1-p$ we
add both directed edges $uv$ and $vu$ to $G$. With probability $p$, we
add one of these two edges picked at random. This represent the fact
that even among the edges corresponding to mutual friendship, some are
not reciprocated. In addition to these edges, we add random directed
edges from $S$ to $V(G)$ and from $V(G)$ to $C$. We underscore that
this models spammers indiscriminately linking to some subset of
\emph{all} nodes, including possibly celebrities and other spammers,
and the converse for inbound links to celebrities. For generating
these edges, we use a simple model of independent coin flips: for each
pair $(u,v)$ where $u\in S$ and $v\in V(G)$, we add $(u,v)$
independently with probability $p_s$. Similarly, for each $(u,v)$
where $u\in V(G)$ and $v\in C$, we add this edge independently with
probability $p_c$. There is no other edge in the graph $G$.

The parameters of the model are as follows: $N$, $|C|$, $|S|$, $p$,
$p_c$, $p_s$, and the parameters of the generation model for the graph
$H$. The algorithm is successful if it gives high $c_v$ scores to
nodes in $C$ (and low $c_v$ to nodes in $V(G)\setminus C$) and high
$s_v$ scores to nodes in $S$ (and low $s_v$ to nodes in $V(G)\setminus
S$).

For the experimental results we present in this paper, we have picked
the following set parameters: $N=2000000$, $|C|=1000$, $|S|=5000$,
$p=0.2$, $p_c = p_s = 0.00025$, and the graph $H$ is a random graph
with expected degree distribution that is a power law with exponent
$0.5$. The average degree in $H$ is 100. These choices are mostly
based on our intuition for typical numbers on a social network. We
have also tried the experiments on several other sets of parameters,
and did not observe any significant change in our conclusions.

\paragraph{LiveJournal Data Set}
Each node in this data set is a LiveJournal profile, and edges
correspond to friendship relationships declared on the profiles.  This
data set is crawled, and contains more than 4.8 million vertices and
660 million edges. LiveJournal users may choose to disallow crawling
of their metadata via the robots.txt mechanism. Any user who did so
was not included in the crawl, with all edges to and from this user
removed from the data set.

\subsection{Convergence speed}
\begin{figure*}
\begin{center}
\subfigure[LJ celebrity score]{%
\includegraphics[width=0.5\textwidth]{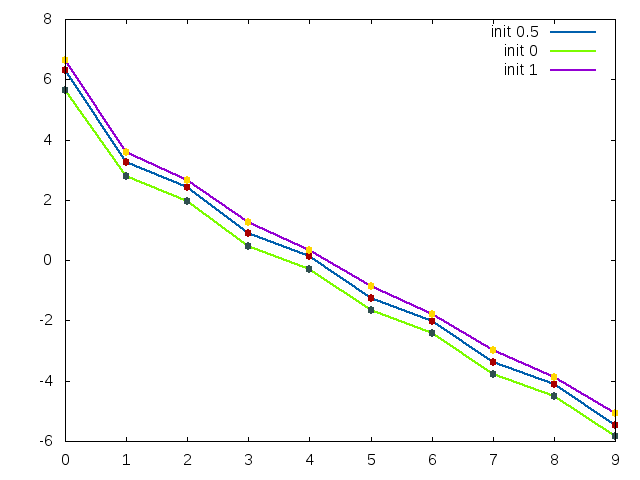}}%
\label{lj_celeb_conv}\hfill
\subfigure[Synthetic graph celebrity score]{%
\includegraphics[width=0.5\textwidth]{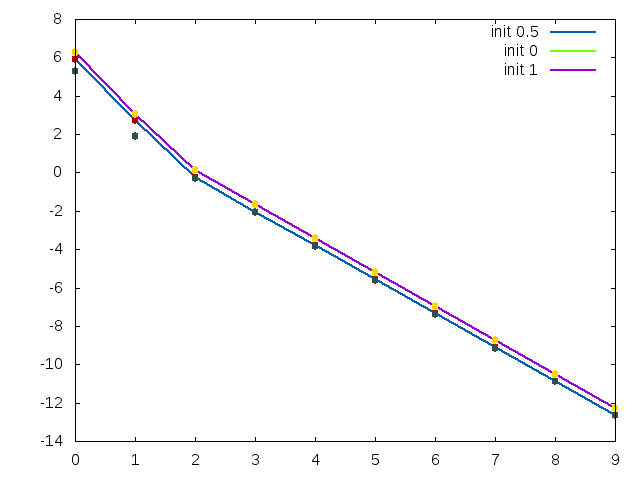}}%
\label{rand_celeb_conv}\\
\subfigure[LJ spammer score]{%
\includegraphics[width=0.5\textwidth]{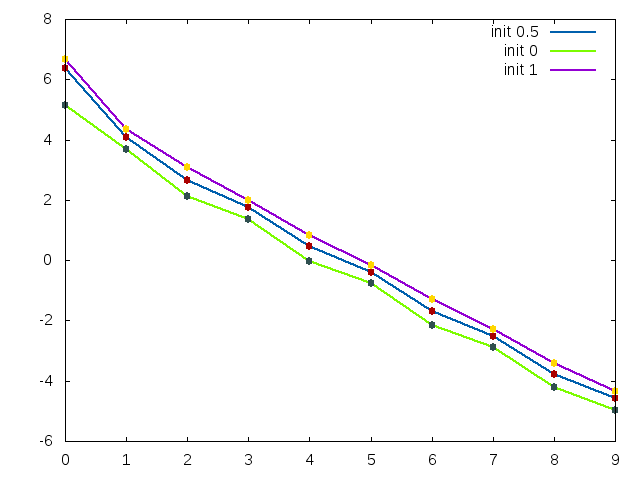}}%
\label{lj_spam_conv}\hfill
\subfigure[Synthetic graph spammer score]{%
\includegraphics[width=0.5\textwidth]{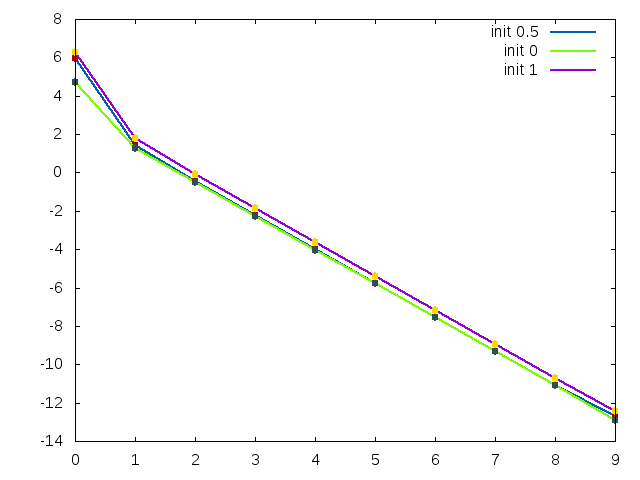}}%
\label{rand_spam_conv}
\end{center}
\caption{Log-scale $l_1$ change in scores for LiveJournal and
  synthetic data, as a function of time}
\label{fig:convergence}
\end{figure*}

Let $\mathbf{c}$ and $\mathbf{s}$ denote the vector of $c_v$'s and
$s_v$'s, respectively.  We can compute the $\ell_1$ distance between
the vector $\mathbf{c}$ computed at the end of iteration $t$, and the
one computed at the end of iteration $t+1$ (and similarly for
$\mathbf{s}$). When both of these values reach zero, it means that the
algorithm has converged to a solution. Therefore, we can use these
values as a measure of the convergence of the algorithm. We plot these
values as a function of $t$ for different data sets and for different
initializations of the scores, to see if and how the initialization
affects convergence speed. The results (in log scale) for the three
data sets are presented in Figure~\ref{fig:convergence}.

The initializations labelled {\tt init 0}, {\tt init 1}, and {\tt init
  0.5} correspond to initializing all scores to zero, all scores to
one, and all scores to $0.5$. We also tried initializing each score to
a random number picked uniformly from $[0,1]$; this initialization
produced results that were essentially indistinguishable from {\tt
  init 0.5} in all data sets.\footnote{Intuitively, this is due to the
  law of large numbers: for most nodes, they have enough neighbors so
  that the sum of the non-celebrity/non-spammer scores of their
  neighbors is essentially the same in {\tt init 0.5} and {\tt init
    rand}.}  As can be seen in the plots, different initializations do
not differ significantly in terms of their convergence rate, although
{\tt init 0.5} often performs marginally better. In all cases, the
convergence seems to be exponentially fast (i.e., the log-scale plot
has an almost constant negative slope)

\subsection{Uniqueness of the solution}
\begin{figure*}
\begin{center}
\subfigure[LJ: celebrity scores]{%
\includegraphics[width=0.5\textwidth]{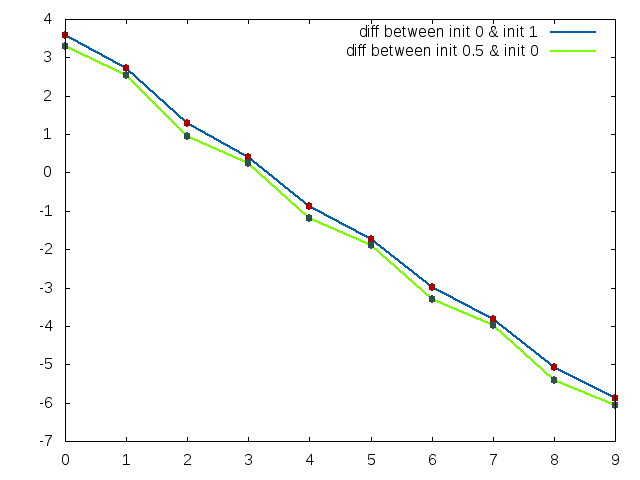}}%
\label{fig:lj_celeb_uniq}\hfill%
\subfigure[Synthetic graph: celebrity scores]{%
\includegraphics[width=0.5\textwidth]{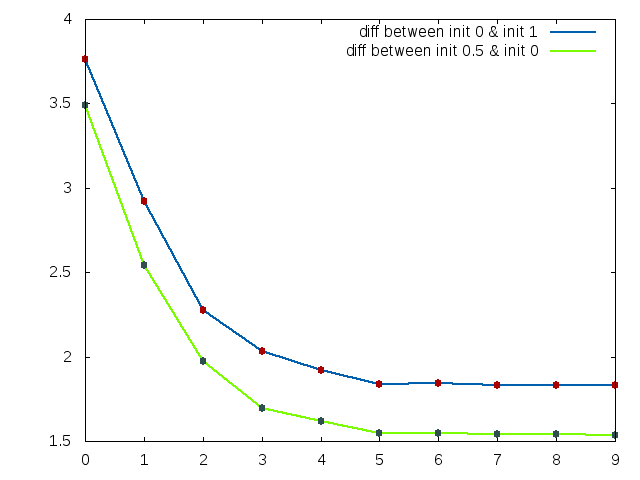}}%
\label{fig:rand_celeb_uniq}\\
\subfigure[LJ: spammer scores]{%
\includegraphics[width=0.5\textwidth]{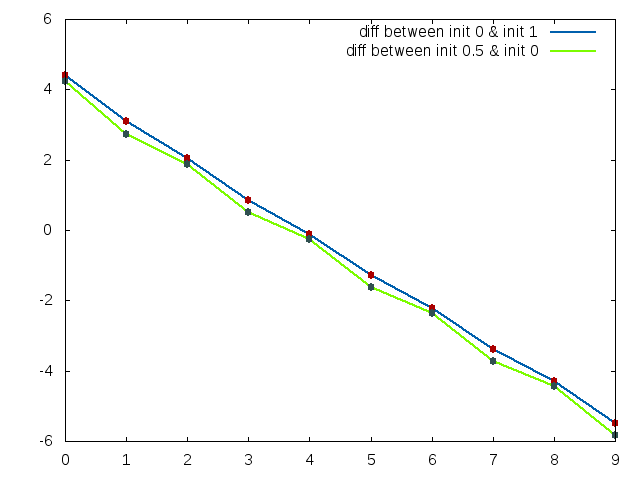}}%
\label{fig:lj_spam_uniq}\hfill%
\subfigure[Synthetic graph: spammer scores]{%
\includegraphics[width=0.5\textwidth]{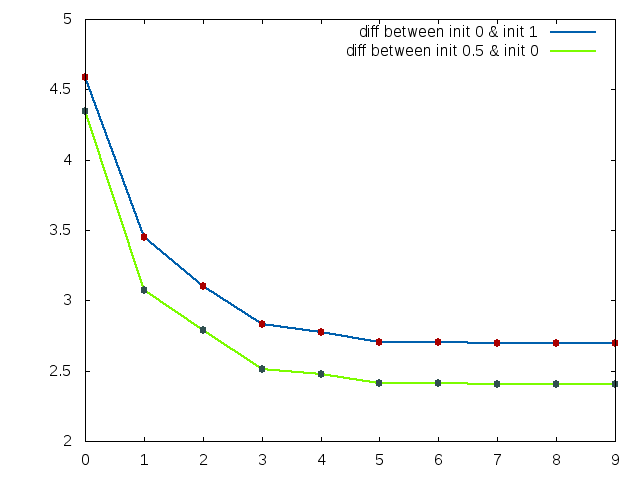}}%
\label{fig:rand_spam_uniq}\\
\end{center}
\caption{Log of $l_1$ difference between scores computed with
  different initializations, as a function of time}
\label{fig:uniqueness}
\end{figure*}

To test whether the scores converge to a single point independent of
the starting point, we plot the $l_1$ distance between the vector
computed by our algorithm starting from different initializations. In
particular, we measure the difference between {\tt init 0} and {\tt
  init 1}, and between {\tt init 0} and {\tt init 0.5}. The graphs,
plotted as functions of $t$ in the log scale, are shown in
Figure~\ref{fig:uniqueness} for the LiveJournal and randomly generated
data sets.


As these graphs show, on the real-world data set after less than $10$
rounds, the solutions computed with different initializations are
virtually identical. In randomly generated instances, even though the
distance between the solutions decrease by about two orders of
magnitude in the first five iterations, they do not converge to
zero. This indicates that randomly generated instances probably
contains small pockets of nodes with non-unique solutions.

\subsection{Solution quality}
\begin{figure}
\begin{center}
\subfigure[Linear-scale spammer
  scores]{\includegraphics[height=0.2\textheight]{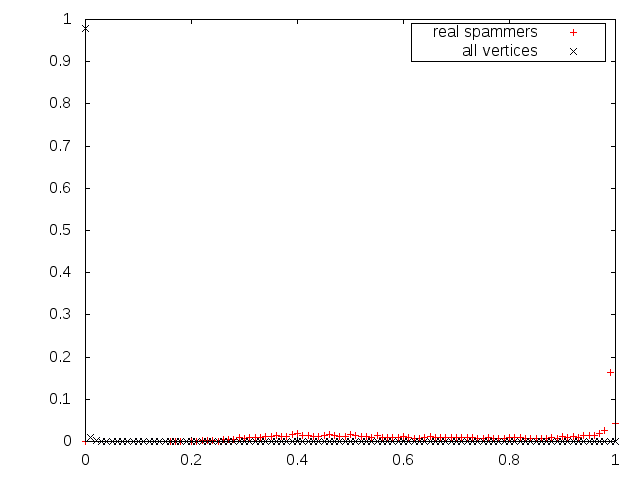}}\\
\subfigure[Log-scale spammer
  scores]{\includegraphics[height=0.2\textheight]{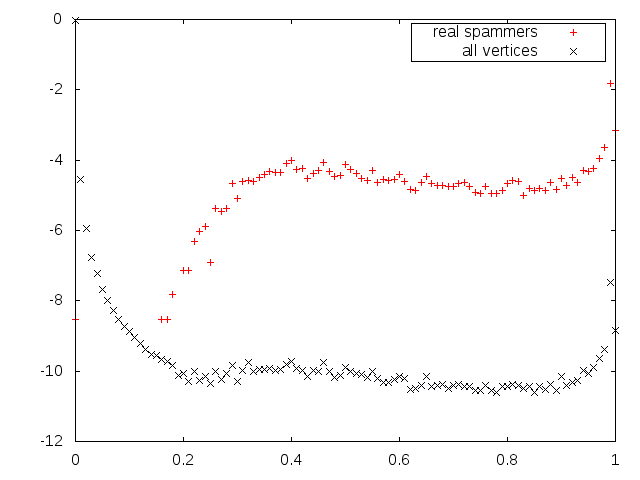}}\\
\subfigure[Linear-scale celebrity
  scores]{\includegraphics[height=0.2\textheight]{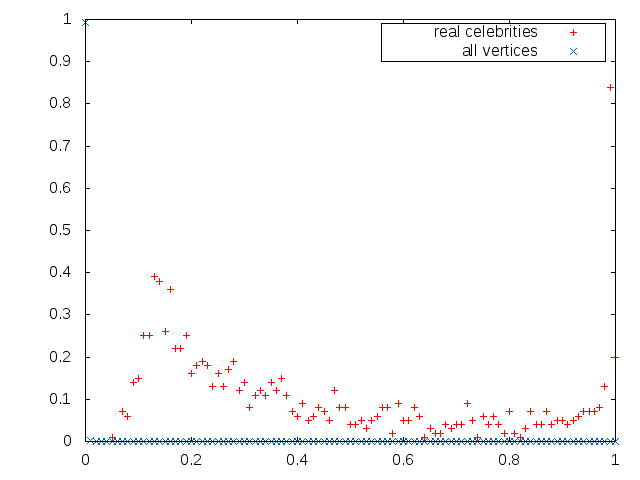}} \\
\subfigure[Log-scale celebrity
  scores]{\includegraphics[height=0.2\textheight]{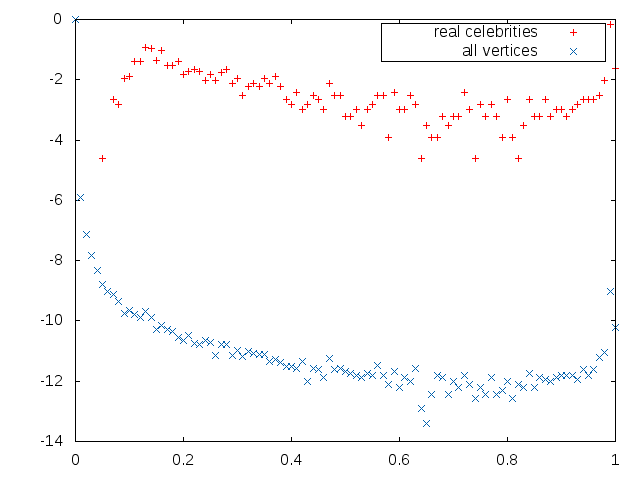}}
\end{center}
\caption{Score distributions (p.d.f.)}
\label{fig:scoredist}
\end{figure}

In this section, we argue that SCRank can recover a significant
portion of celebrities and spammers. To show this experimentally, we
use randomly generated graphs with the sets $C$ and $S$ in the random
generation process as the {\em hidden ground truth}. The algorithm is
successful if it assigns high celebrity scores to nodes in $C$ and
high spammer score to nodes in $S$. Figure~\ref{fig:scoredist} shows
the distribution of celebrity and spammer scores, comparing,
respectively, all vertices versus vertices in $C$; and all vertices
versus vertices in $S$. The score distributions on these synthetic
inputs are almost completely bimodal, with both celebrity and spammer
scores of generic vertices being strongly concentrated around zero.
To better observe the difference between the two distributions, we
also show plots of the distribution densities with logarithmic
$y$-axes.

\begin{figure}
\begin{center}
\includegraphics[height=0.25\textheight]{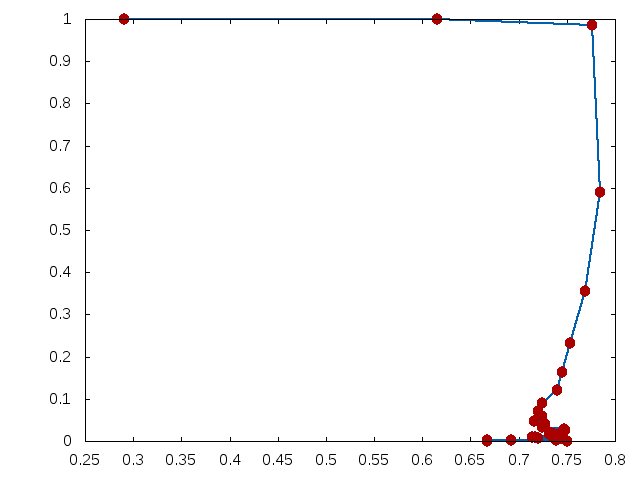}
\end{center}
\caption{Precision-recall graph for the detection of spammers in a randomly generated data set.}
\label{fig:precrecall}
\end{figure}

We can also study the precision-recall tradeoff of the output of the
algorithm.  We plot the precision of the algorithm (which we define as
the percentage of users with celebrity/spammer score more than $0.5$
who are in $C$/$S$, respectively) against recall (defined as the
percentage of nodes in $C$/$S$ for which we compute a celebrity or
spammer score, respectively, of more than $0.5$). By adjusting the
parameters of the model, we get a tradeoff between precision and
recall that is plotted in Figure~\ref{fig:precrecall}.

\section{Generalization}
\label{sec:gen}

The potential function argument used in Section~\ref{sec:conv} to
guarantee SCRank's convergence can be generalized to a much broader
class of iterative algorithms, which we expect will be of independent
interest. In particular, we will show that the same argument applies
to any iteration that simulates \emph{best-response dynamics} in a
game where players have bounded real-valued strategies, and
whose utilities are strictly monotic, continuously
differentiable per-variable functions which depend only on a linear
combination of the others' strategies, with \emph{symmetric} linear
combination weights.

\newcommand{\RR}{\ensuremath{\mathbb{R}}}
\newcommand{\ZZ}{\ensuremath{\mathbb{Z}}}

Before we define this formally, let us observe how this describes the
SCRank algorithm. SCRank uses $2n$ variables --- two ``players'' per
SCRank agent. The convergence argument can be rephrased to ignore the
fact the variables are arranged in pairs in the original setup. The
updates in lines 6--11 of the algorithm are equivalent to 
the $c_i$ players making best-response moves one at a time, then
the $s_i$ players taking their turns.
The utility/update functions for the $c_v$s and $s_u$s both depend only
on a linear combination of other variables: $F_{c_v} (L =
\sum_{u|(u,v)\in A} -s_u) = F_c(\operatorname{indeg} v + L)$, and
similarly for $F_{s_u}$.  For every $(u,v)\in A$, $c_v$'s update
function input will include $s_u$ with weight $-1$, and $s_u$'s update
function input will include $c_v$ with weight $-1$. This meets the
``symmetricity'' condition --- that the matrix $W$ of variable weights
in the update function inputs must be symmetric. In SCRank, $W_{c_v
  s_u} = W_{s_u c_v} = -1$, and is 0 elsewhere.

\newcommand{\MUSLI}{M\"USLI}

Formally, we define a general \emph{Monotonic \"Update on Symmetric
  Linear combinations Iteration} (\MUSLI) system as:

\begin{itemize}
\item Real-valued variables $x_1,\ldots,x_n$ with $x_i \in [a_i, b_i]$
\item A symmetric weight matrix $W$ with 0s on the
  diagonal.
\item For each variable, a strictly increasing, continously
  differentiable update function $F_i:\RR \to \RR$ which takes as
  input only $(Wx)_i$, the linear combination of the $x_i$s weighted by $W$'s
  $i$th row. $F_i$ must preserve the bounds on $x_i$, i.e. $a_i \leq
  F_i((W\vec{x})_i) \leq b_i$ whenever $x_j\in [a_j, b_j]$)
\item An activation sequence $A:\ZZ_{\geq 0} \to \{1,\ldots,n\}$
  determining, for each iteration $t \geq 0$, the unique variable
  $x_{A(t)}$ that gets updated to $F_{A(t)} ((Wx)_{A(t)})$.
\end{itemize}

The proof of Theorem~\ref{thm:conv} generalizes to show:

\begin{theorem}
The state of a \MUSLI\ system, $\vec{x}$, converges to a fixed point
under its iteration.
\end{theorem}

\begin{proof}
The argument is very similar, relying just on a generalization of the
potential function. As above, the strictly increasing,
continuously differentiable $F_i$s have well-defined strictly
increasing inverses $F_i^{-1}$, which lets us define $G_i(z)$ and the potential
function $P(\vec{x})$ as:

\begin{align*}
G_i(z)
& = \int_0^z F_i^{-1}(t) dt \quad; \quad
P(\vec{x})
 = \sum_i G_i(x_i) - \frac{1}{2} \vec{x}^T W \vec{x}
\end{align*}

This yields the partials:

$$\frac{\partial P}{\partial x_i} = -(W\vec{x})_i + F_i^{-1}
(x_i)$$.

For $x^*_i = F_i((W\vec{x})_i)$, this is zero, and, since $W_{ii}
= 0$, the first term is constant relative to $x_i$, and the
monotonicity of $F_i^{-1}$ guarantees that updating $x_i$ to $x^*_i$
can't increase $P(\vec{x})$.

\newcommand{\xOLD}{x^{\mathrm{old}}}

As before, $0 < \frac{dF_i(t)}{dt} \leq \alpha_i$ for some $\alpha_i$,
and $\frac{dF^{-1}_i(t)}{dt} \geq 1/\alpha_i$. An iteration that
starts at state $\xOLD$ and updates $x_i$ from $\xOLD_i$ to, WLOG,
a lower value $x^*_i < \xOLD_i$, changing it by $\xOLD_i - x^*_i \geq
\varepsilon$, will have, for all $t\in [x^*_i, \xOLD_i]$:

\begin{align*}
F^{-1}_i(t) & > F^{-1}_i(x^*_i) + (t-x^*_i)/\alpha_i \\
& = (W\xOLD)_i + (t - x^*_i) / \alpha_i \\
\frac{\partial P}{\partial x_i} (\xOLD_{-i}, t) & \geq (t - x^*_i) / \alpha_i \\
P(\xOLD) - P(\xOLD_{-i}, x^*_i) & \geq \int_{\xOLD_i}^{x^*_i} \frac{t -
x^*_i}{\alpha_i} dt \geq \frac{\varepsilon^2}{2\alpha_i}
\end{align*}

Since $\vec{x}$ remains within the compact set defined by $x_i \in [a_i,
  b_i]$, $P(\vec{x})$ is bounded, and, since it decreases at each step
of the iteration, there is, by the same argument as above, for any
$\delta > 0$, a step $k^*$ such that the update won't change $x_i$ by
more than $\sqrt{2\alpha_i \delta}$.
\end{proof}

Note that the proof doesn't even require that each variable be
``activated'' infinitely often, but we expect most practical uses of
this result will require that each $x_i$ be updated infinitely often
for convergence to a relevant value, or more often than some threshold
for stronger convergence bounds.

To demonstrate the breadth of these systems, we now give a couple of
examples.

\subsection{Example: Graph connectivity}

As a trivial example of another algorithmic task solvable via a
\MUSLI\ best-response iteration, consider the question of (undirected)
graph reachability. If the graph's adjacency matrix is used as
weights, with nodes as players, using starting state 0 for all players
except the origin, iterating updates of sigmoid $F_i(\sum_{N(i)} x_i)$
that approximates a step function at $x = 1$ will clearly converge
rapidly to a state where all nodes reachable from the origin are $1$.

\subsection{Example: Influence games}

In SCRank and the above example, \MUSLI\ systems are used as
algorithms to compute a fixed point of interest. We note that the
one-at-a-time update dynamics and the $W_{ii} = 0$ constraint mean
that \MUSLI\ iterations can also be interpreted as classical
best-response dynamics in games, immediately yielding:

\begin{corollary}
A game whose best-response dynamics form a \MUSLI\ system (i.e. an
$n$-player game with bounded real-valued strategies and strictly
increasing, continuously differentiable best-response functions that
depend only on a linear combination of the other players, with
symmetric weights) is a potential game \cite{potentialgames}, with the
above potential function, and is guaranteed to converge.
\end{corollary}

This class of games is fairly broad, including, for instance:

\noindent The \textbf{party affiliation game}. In the well-studied
party affiliation game \cite{FPT04}, agents pick ``political parties''
$-1$ and $1$ based on the weighted sum of their friends' and enemies'
affiliations: a player tries to be in the same party as her friends
and in a different party than her enemies. Allowing fractional
strategies and softening the best-response function from the original
step function $F_i(\vec{x}) = \operatorname{sgn} \sum_j x_i x_j
w_{ij}$ to a sigmoid that approximates it produces a game whose
best-response dynamics are a \MUSLI\ system. The above argument
guarantees a potential function and convergence. We note that this is
quite natural, since our potential function argument is an extension
of the max cut game potential argument that underlies the analysis of
the party affiliation game.

\noindent The \textbf{symmetrical technology diffusion game}. Consider a
social network where agents are deciding, e.g., between 2 technologies
with a network effect such as cellular providers where a user benefits
from having more friends use the same technology. In the US cellular
market, this corresponds to free phone calls to people on the same
network, and heavy charges for calls to people on another network
beyond a fixed monthly limit. Let weight $W_{ij}$ indicate how many
minutes $i$ and $j$ expect to talk on the phone per month, $0$ and $1$
represent the provider choices, and $\vec{x} \in [0,1]^n$ be the
current fractional provider choices, optionally considered as
probabilities. A natural best-response function for $i$ is to use
$(W\vec{x})_i$, the expected number of minutes she will spend talking
to people using provider 1 (assuming minutes and provider choices are
independent), as an input to a sigmoid that is a soft step function at
or near the maximum number of free calling minutes for users of
provider 0 when calling users of provider 1. Assuming all phone calls are
2-way, the best-response dynamics constitute a \MUSLI\ system,
immediately demonstrating that the game is a potential game
and guaranteeing convergence.

\section{Conclusion}
In this paper, we presented a framework for iterative algorithms for
giving scores to nodes defined recursively in terms of the scores of
their neighbors, with a focus with one application in which such a
recursive definition comes quite naturally: computing celebrity and
spammer scores on a directed social network. We theoretically proved 
that under a mild symmetry and monotonicity assumption, there is 
a potential function that decreases in every iteration of the
iterative algorithm, and therefore, the iterative algorithm always
converges to an approximate equilibrium. In the case of
celebrity/spammer scoring, we experimentally showed that this
convergence is extremely fast, the convergence point is unique, and
when applied on randomly generated data with a built-in ground truth,
it provides a good approximation to the ground truth. 

In addition to the obvious application of finding celebrities and
link-spammers in online directed social networks, we believe that 
our potential function framework has the potential to be quite useful
in theoretical analysis of iterative algorithms on social
networks. Iterative algorithms such as belief propagation are
notoriously hard to analyze theoretically, despite widespread
practical use.

The obvious open directions are to find other applications or
generalizations of our framework, or prove a theoretical bound on the
convergence speed of the algorithm that is close to the practical
observation.

\bibliography{scrank}{}
\bibliographystyle{abbrv}

\end{document}